\documentclass[journal]{IEEEtran}
\usepackage{cite}
\usepackage{amsmath,amssymb,amsfonts}
\usepackage{algorithmic}
\usepackage{graphicx}
\usepackage{textcomp}
\usepackage{xcolor}
\usepackage[numbers,sort&compress]{natbib}
\usepackage{hyperref}
\usepackage{soul}
\usepackage{multirow}

\ifCLASSINFOpdf
\else
\fi
\begin{document}
%
\title{A Topic-Attentive Transformer-based Model For Multimodal Depression Detection}
%
%
%

\author{Yanrong~Guo,
        Chenyang~Zhu,
        Shijie~Hao,
        and~Richang~Hong,~\IEEEmembership{Senior Member,~IEEE}
\thanks{Y. Guo, C. Zhu, S. Hao, and R. Hong are all with Key Laboratory of Knowledge Engineering with Big Data (Hefei University of technology), Ministry of Education and School of Computer Science and Information Engineering, Hefei University of Technology (HFUT), 230009 China e-mail: yrguo@hfut.edu.cn.}
}

\maketitle

\begin{abstract}

Depression is one of the most common mental disorders, which imposes heavy negative impacts on one's daily life. Diagnosing depression based on the interview is usually in the form of questions and answers. In this process, the audio signals and their text transcripts of a subject are correlated to depression cues and easily recorded. Therefore, it is feasible to build an Automatic Depression Detection (ADD) model based on the data of these modalities in practice. However, there are two major challenges that should be addressed for constructing an effective ADD model. The first challenge is the organization of the textual and audio data, which can be of various contents and lengths for different subjects. The second challenge is the lack of training samples due to the privacy concern. Targeting to these two challenges, we propose the TOpic ATtentive transformer-based ADD model, abbreviated as TOAT. To address the first challenge, in the TOAT model, topic is taken as the basic unit of the textual and audio data according to the question-answer form in a typical interviewing process. Based on that, a topic attention module is designed to learn the importance of of each topic, which helps the model better retrieve the depressed samples. To solve the issue of data scarcity, we introduce large pre-trained models, and the fine-tuning strategy is adopted based on the small-scale ADD training data.
We also design a two-branch architecture with a late-fusion strategy for building the TOAT model, in which the textual and audio data are encoded independently.
We evaluate our model on the multimodal DAIC-WOZ dataset specifically designed for the ADD task. Experimental results show the superiority of our method. More importantly, the ablation studies demonstrate the effectiveness of the key elements in the TOAT model.

\end{abstract}

\begin{IEEEkeywords}
Multimodal depression detection, transformer, topic attention
\end{IEEEkeywords}

%
\IEEEpeerreviewmaketitle

\section{Introduction}
\IEEEPARstart{D}{epression} is one of the most common mental disorders in nowadays society. Patients suffering from depression may have difficulty in focusing attention, keeping a good mood, or have suicidal tendency in extreme conditions \cite{2019Providing}. According to \cite{dep_refer1}4.4\% people of the world are now suffering from depression, and the prevalence is still on the rise. The good news is that depression can be effectively alleviated or be cured with an in-time diagnosis. However, unlike other physical disease, depression symptoms are usually obscure, making the depression detection difficult. Currently, the diagnosis of depression relies commonly on subjective rating scales, such as the Beck Depression Inventory (BDI) and the Eight-item Patient Health Questionnaire depression scale (PHQ-8) \cite{kroenke2009the_14}. The accurate diagnosis based on these kinds of information usually depends on the participation of experienced psychologists, which can be labor-intensive for many occasions. Thus, the urgent demand for a practical and accurate method for detecting depression gives rise to a new task based on machine learning techniques: automatic depression detection (ADD) \cite{wang2008automated_1}. ADD systems typically work through recording and analyzing data from interviewees, such as speech or facial expression, which is more objective and requires fewer expert interventions.

Prior ADD works \cite{2021ta1, hanai2018detecting_6, 2019Detecting} have shown that information from one modality is limited, while different modalities such as audio, text, and video are complementary and have the potential of improving the ADD performance. Therefore, multimodal depression detection has been attracting more and more attention in recent years. As the cornerstone, the DAIC-WOZ dataset \cite{daic} is designed and built as a multimodal depression dataset, which becomes the most popular platform for conducting the ADD research. During the data collection, an AI is adopted to ask participants fundamental questions and collects audio, video, and depth sensor recordings during the interviews. Among the modalities, we can see that speech and its text transcripts are easy to obtain in daily life. Therefore, it is more practical to build an ADD system that is based on the audio and textual modality data. For example, the data can be simply collected through phone calls. In this context, we plan to construct an effective ADD model by fully exploring the data from these two modalities.

To achieve this goal, we have two major challenges to address. The first one is the reasonable organization of the speech data. For different subjects, their interviewing contents may be diverse. For example, the question list can be different among subjects, and the answers of a same question can be varied from one to another in terms of the contents and the duration time. In this context, it is important to organize the raw data into a suitable form that facilitates the full exploration of ADD-specific features. The second challenge is the data scarcity due to the privacy concern. Despite of the prevalence of depression, it is not easy to collect a large-scale dataset for ADD research, as most people are less willing to authorize the use of their data even if the research ethics can be guaranteed. This issue poses difficulty to the learning-based ADD task, which usually requires sufficient training data for fitting the complex mapping between the multimodal data and the semantic label.

In our research, we propose the TOpic ATtentive transformer-based ADD model to address the above challenges. First, we organize the audio and textual data by taking topic as the basic unit. In a typical data collection scenario, a subject answers pre-defined questions one by one during the interview. Therefore, the organization of the original data is to take each question as a single topic, which is the basic unit for building the ADD model. The consideration behind this idea are two-fold. On the one hand, the topic is at the mesoscopic level that stays between the sentence level and the subject level. A topic contains semantic information that is more complete than a single sentence. In the meanwhile, the information from multiple topics of a subject can be more comprehensive for evaluating his/her mental state. On the other hand, more importantly, we can introduce the topic attention mechanism, which enables our model to estimate the importance of each topic for accurately inferring the subject's mental state.

Second, we relieve the data scarcity issue by introducing and fine-tuning pre-trained models. Recently, transformer-based \cite{transformer} models based on self-supervised learning (SSL) have achieved great success in natural language processing \cite{devlin2018bert, yang2019xlnet, raffel2020exploring}, computer vision \cite{zhou2021ibot, bao2021beit, he2021masked}, and automatic speech recognition \cite{wav2vec2, baevski2020vq, schneider2019wav2vec}. These SSL models are pre-trained on a large scale of unlabelled datasets for learning feature representations, and they can be effectively applied on downstream tasks. Several pioneering works \cite{devlin2018bert} have demonstrated the superiority of fine-tuning these SSL models on downstream tasks, which sheds lights on solving the data scarcity issue in our application.

\begin{figure*}[htpb]
		\centering
		\includegraphics[width=\textwidth]{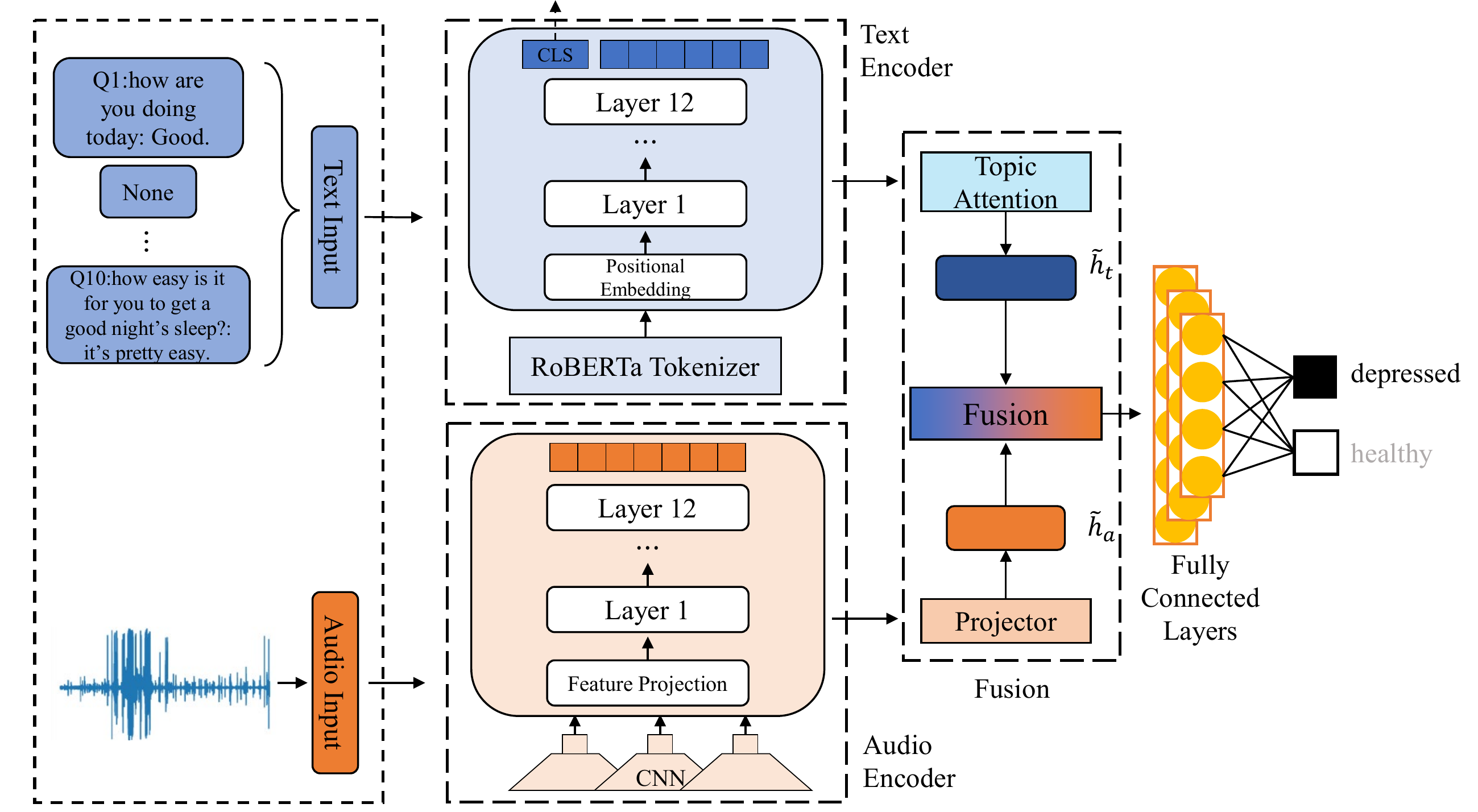}
		\caption{The architecture of our proposed model.}
		\label{framework}
	\end{figure*}

Based on the above considerations, we design the proposed TOAT model as a two-branch structure (\autoref{framework}), in which two pre-trained models are introduced. As for the textual modality, we introduce and fine-tune the RoBERTa model \cite{roberta} to encode the texts of each topic into the depression-related features. As for the audio modality, we introduce and fine-tune wav2vec 2.0 \cite{wav2vec2} to encode the raw audio signals into the features that can be used for the ADD task. The complementary of the two modalities is utilized by gathering the two branches in a simple fusion module, which also helps to relieve the data insufficiency issue to some extent.

 The technical contributions of this research can be summarized as follows:
\begin{itemize}
    \item We propose a novel architecture for multimodal depression detection. The introduction of pre-trained models, as well as the fusion mechanism, addresses the data scarcity issue in the ADD task.

    \item We design the topic attention module to learn the importance of the questions for a participant during the interview. From the experiments, we empirically find that the attention mechanism boosts the system by enhancing its ability of retrieving more positive samples from the participants.

\end{itemize}

The rest of the paper is organized as follows. Section II introduces the related works. Section III introduce the overall framework of our TOAT model and its details. In Section IV, we provide the experimental results and their analysis. Section V finally concludes the paper.


	\section{Related Work}

	 Multimodal depression detection infers one's mental state by jointly exploring data of multiple modalities, such as visual, acoustic, or textual modalities. In this section, we discuss the multimodal fusion and topic-based methods in the depression detection task.
	
	 Fusion-based modeling is a common technical roadmap, in which intelligible and appropriate features are firstly extracted from different modalities, and the fusion of these features is then conducted at different stages of the method. Valstar et al. \cite{2016avec_3} propose a baseline fusion method at the inference level, in which a logic AND operation for the audio-based inference and the video-based inference is made. Yang et al. \cite{decision_tree} propose a gender-specific decision tree that utilizes three modalities, of which the statistical features cooperate with each other along the built trees. Qureshi et al. \cite{oureshi2021gender} propose the gender-aware model, in which the multi-modality features are concatenated, and a fusing vector is learned to assign different weights for them. Ray et al. \cite{ray2019multi_7} propose a novel framework that builds the two-level fusion mechanism. The first-level fusion is conducted within each modality, and the second-level fusion is conducted among the different modalities. As pre-trained models have become more and more effective to extract features, Makiuchi et al. \cite{makiuchi2019multimodal} introduce BERT to extract text-modality features. For the audio modality, deep spectrum features are extracted by using a pre-trained VGG-16 network \cite{vgg}, and a gated CNN followed by a LSTM layer \cite{lstm}. In the fusion part, these two kinds of features are concatenated directly. From the above works, we can see that the late-fusion strategy is mainly adopted in the multimodal depression detection task, in which the main focus is either on the effective feature extraction, or the accurate fusion-weight learning. Despite the subtle modeling, a main challenge for these methods lies in the small dataset, which tends to bring in the over-fitting issue, especially for the models with large solution space. It is noted that the well pre-trained models have the potential of solving this issue, in which the fine-tuning strategy can be applied based on the small-scale dataset for depression detection.

	 From another perspective, the multimodal depression detection models are built upon the collected data at different scales. Specifically, scale means the observation range of the multi-modality data. For example, Rohanian et al. \cite{rohanian2019detecting} perform fusion after each word in an utterance, in which fusion gates are adopted to control the contributions from audio and video modalities. The observation range of this model is at the word level, which is very local. In contrast, many above-mentioned methods \cite{2016avec_3}\cite{oureshi2021gender} \cite{ray2019multi_7} directly extract and fuse features at the global level, i.e. the whole input data is taken as the input of the feature extractor. Apart from the methods based on the microscopic scale or the macroscopic scale, modeling the multi-modality data at the mesoscopic scale is also useful for the depression detection task. As the interview for depression detection typically has a question-answer form, the structured data facilitates the construction of a topic-based model. It is noted that it is different from the traditional topic model \cite{hong2010empirical} aiming at discovering topics from documents. Instead, the topic-based models here treat a question as a topic directly. Gong et al. \cite{gong2017topic_5} propose a topic-modeling-based method to perform context-aware analysis. They concatenate visual, textual and audio features based on topics to build input data. The research in \cite{toto2021audibert} also considers the topic as the basic unit during the modeling.
	
	 Our model extracts the features of different modalities at the topic level as well, and simply fuses them before sending them into a linear classifier.
	 As for the topic-based modeling, our model is different from \cite{gong2017topic_5} and  \cite{toto2021audibert}, as we construct a topic attention model to enable the model to learn the topic importance. Of note, in \cite{toto2021audibert}, the whole dataset containing multiple topics is re-organized by splitting it into topic-wise subsets, while our model is still built upon the whole dataset. Therefore, the difference of the data organization makes the two methods fundamentally different in terms of the model construction and performance evaluation.
	
	\section{Method}
	
	\subsection{Overview}
	\autoref{framework} shows the framework of our TOAT model. The model input includes the textual data and the corresponding audio data of $N$ topics, which is prepared for the two branches in our model. One branch aims to explore the depression-related semantic information from the textual modality data. The other branch aims to explore the acoustic information that is correlated with depression from the audio modality data. Then, the learned features from the two branches are directly combined and sent into a linear classifier. In this way, the depression detection is formulated as a binary classification problem (being positive or negative).

	\subsection{Learning Textual Modality}
     As for the textual branch, we detail the process ahead of the topic attention. It has been proved that fine-tuning a BERT-like model to obtain textual features can be effective for the depression detection task \cite{toto2021audibert}. Besides, considering the data deficiency, a complex model can be easily overfit if trained from scratch. Therefore, we fine-tune a pre-trained improved version of BERT, i.e. RoBERTa, to extract textual features.

     Firstly, for every participant, we build a group that consists of $N$ frequently-asked questions and their replies. We represent each topic as $S_i$ ($i=1,...,N$), which contains a topic index $i$, the corresponding question and reply $S_i = \{q_i,r_i\}$. Therefore, the textual data of $N$ topics is represented as $X_{t}=\{S_1, S_2, \cdots S_{N}\}$.

     To extract semantic information, we introduce RoBERTa \cite{roberta} as the backbone, which is an upgraded BERT via training on larger datasets containing longer sentences with larger batches. Besides, it removes the next sentence prediction objective in BERT \cite{devlin2018bert}. RoBERTa has two versions, i.e. RoBERTa-large and RoBERTa-base, of which the architecture contains 24 transformer blocks and 12 transformer blocks, respectively. In our application, we simply choose the RoBERTa-base version for the balance between performance and time complexity.

    $\{S_i\}$ are sent into RoBERTa one by one, in which the tokenizer encodes them as
    \begin{equation}
    tokenizer(S_i)=\{[CLS], s_{i,1}, s_{i,2},\cdots,s_{i,l},[SEP]\}
    \end{equation}
    where $s$ is a byte-level token. $CLS$ and $SEP$ tokens are inserted at the beginning and the end of $S_i$. The tokens are further sent into RoBERTa-base's main body, which contains 12 transformer layers and has the  hidden dimension of $D=768$. As we are doing a classification task, the $CLS$ token is taken as the output of RoBERTa, which is denoted as $h_{t_i}\in\mathbb{R}^{1\times D}$. For all the $N$ topics, we concatenate their $CLS$ heads $H_t=[h_{t1}; h_{t2};\cdots h_{t{N}}]$.

    Of note, it is common that not all the questions are asked for each participants in our applications. To solve this practical issue, the missing ones are marked as $None$. For those $None$ parts, we directly fill the corresponding $h_{ti}$ with numbers approaching $-inf$ (-1e9 in our implementation). To an extent, it can be regarded as a mask operation.

   	\subsection{Topic Attention Module}
	
	\begin{figure}[htpb]
		\centering
		\includegraphics[width=\linewidth]{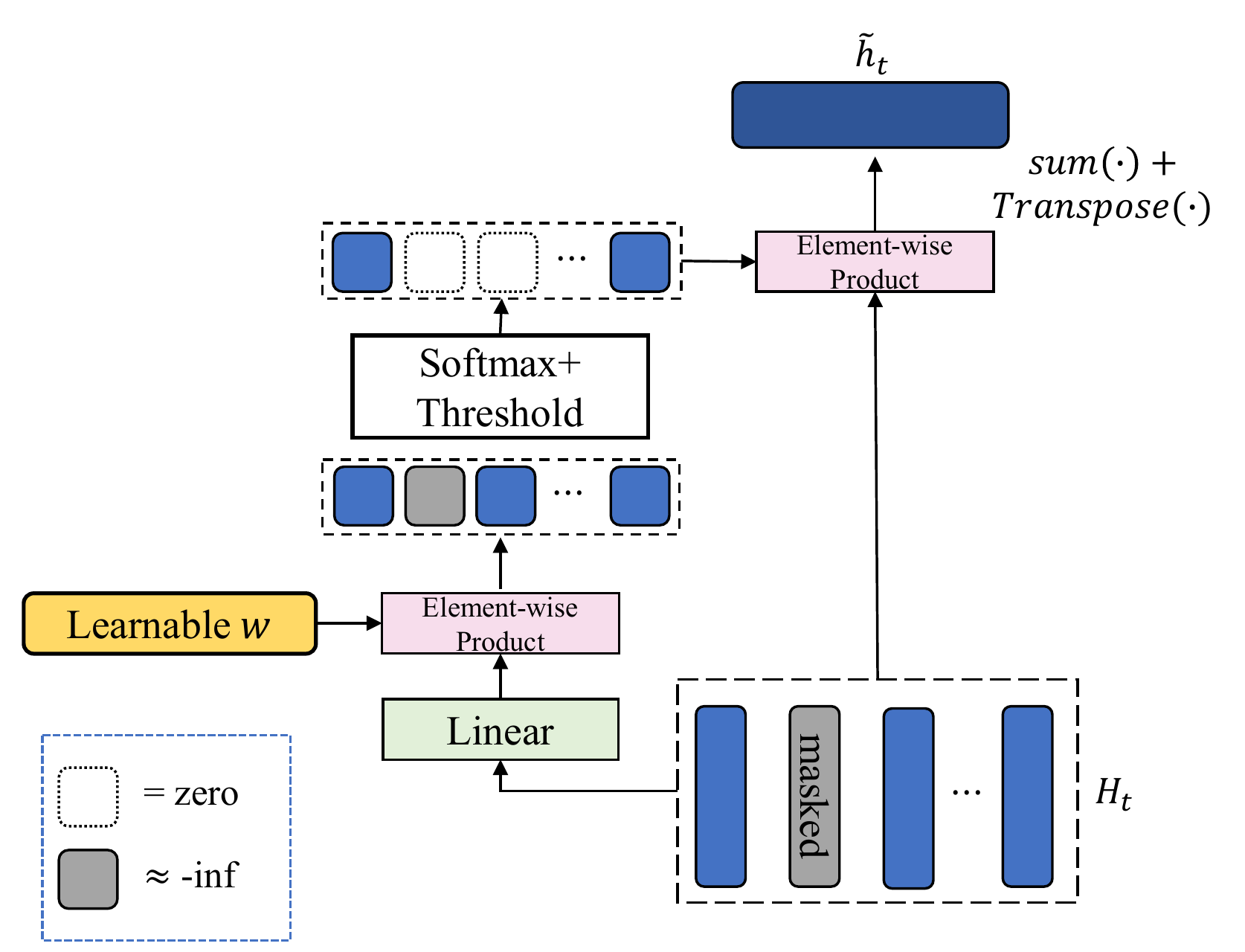}
		\caption{The architecture of the topic attention.}
		\label{tam}
	\end{figure}
	
	 It is plausible to assume that the topics are of varying significance for the depression detection task. For example, it seems that the question like "How easy is it for you to get a good night's sleep?" is more helpful than "What'd you study at school?", as reply of the former question possibly contains important information about the mental state of the participant. Based on this consideration, we introduce the topic attention mechanism, and design the topic attention module to learn the importance of all the adopted topics. As shown in \autoref{tam}, we feed the learned representation $H_t\in \mathbb{R}^{D\times N}$ into a linear layer to reduce its dimension into $1\times N$. Aiming to help the linear layer adjust the topic score, an element-wise product is imposed by a learnable parameter $w\in\mathbb{R}^{1\times N}$. This process can be expressed as:
	\begin{equation}
	    g=w\circ \mathrm{Linear}(H_{t})
	\end{equation}
	Then, we apply a softmax layer $g^*_i = \frac{exp(g_i)}{\sum_{j=1}^{N}exp(g_j)}$ to normalize the topic scores. Of note, the masked vectors do not have influence on this process as $exp(-inf)=0$. Based on the topic score estimation, the model is able to learn the importance of each topic with respect to the depression detection.
	
	We set a threshold $\alpha=1/N$ to penalize these normalized topic scores, and obtain the refined scores $\tilde{g}$ as:
    \begin{equation}
    \tilde{g_i}=
    \begin{cases}
    0 & \text{ if } g_i^*<\alpha  \\
    g_i^* & \text{ if } g_i^*>=\alpha
    \end{cases}
    \end{equation}
    This non-linear threshold operation intentionally highlights the important topics while ignoring the less important ones. Finally, we multiply scores $\tilde{g}_i$ and matrix $H_t$ followed by a sum operation to obtain the weighted representations $\tilde{h}_t$:
    \begin{equation}
        \tilde{h}_t=\sum_{i=1}^{N}\tilde{g}_i H_{ti}
    \end{equation}

	\subsection{Learning Audio Modality}
    In our application, we aim to extract depression-related features from audio data, such as discriminating cues in pronunciation and intonation, to assist the textual information. To this end, we introduce the pre-trained Wav2vec 2.0 model as the backbone of our second branch, which is trained on the Librispeech dataset \cite{librispeech} and achieves the SOTA performance on several audio-related tasks. Different from some previous works \cite{dai2021improving, niu2021hcag} using low-level acoustic features as inputs, we directly use the raw speech signals of the replies from the participants as the input of the pre-trained model.

    The architecture of Wav2vec 2.0 consists of a feature encoder composed of several CNNs, and a transformer module that is similar to RoBERTa. The difference between them is that Wav2vec 2.0 replaces the fixed position embedding with a convolutional layer that works as relative positional embedding. As there is no $CLS$ head for Wav2vec 2.0, we use a projector layer instead, which reduces the dimension with a linear layer, and calculates the average along the timeline:
    \begin{equation}
    \begin{aligned}
        \Bar{H}_a = \mathrm{Linear}(H_a) \\
        \tilde{h}_{a}= \frac{1}{L}\sum_{i=0}^{L}\Bar{H}_{a,i}
    \end{aligned}
    \end{equation}
    Here $H_a\in \mathbb{R}^{L\times D}$ is the output of the last transformer layer, $L$ is the timestep. In our implement, $D_a$ is set as 256. $\tilde{h}_a$ is the obtained feature from the audio branch. $a$ is the abbreviation of audio.

    In the implementation, we ignore the quantization module in a typical Wav2vec 2.0 model, which is not applicable in our application. In addition, considering the computational efficiency, we extract the features by choosing only a part of interviewees' replies, and truncate them into a fixed size.

	\subsection{Multi-modal Fusion}
	In the fusion stage, to demonstrate the usefulness of the feature extraction stage, we directly concatenate the two features $\tilde{X}=concat(\tilde{h}_t,\tilde{h}_a)$, followed by layer normalization \cite{ba2016layer} and a dropout layer, aiming to mitigate the gap between the two kinds of features. Finally, a linear classification layer is adopted to obtain the prediction score $y$.
	
	\subsection{Optimization Objectives}
	To measure the difference between the ground-truth and prediction, we choose the cross entropy loss as the optimization objective for the classification task. The loss function is shown as follows:
	\begin{equation}
	    L=\frac{1}{N}\sum_{i=1}^{N}-y^{(i)}\mathrm{log}y^{(i)}-(1-\hat{y}^{(i)})\mathrm{log}(1-\hat{y}^{(i)})
	\end{equation}
	where $y$ is the prediction and $\hat{y}$ is the ground-truth label.
	
	\section{Experiments}
	

    \begin{table*}[]
    \centering
    \label{testset}
    \caption{Comparsion with several methods based on the DAIC-WOZ validation and test set. 'A', 'V', 'T' denotes the audio, visual and textual modality, respectively. Best results are highlighted.}
    \begin{tabular}{cc|cccc|cccc}
    \hline
                                       &          & \multicolumn{4}{c|}{Validation Set}          & \multicolumn{4}{c}{Test Set}                                      \\ \cline{1-2} \cline{4-5} \cline{8-9}
    \multicolumn{1}{c|}{Model}         & Modality & Accuracy & Recall & Precision & F1-score & Accuracy      & Recall        & Precision     & F1-score      \\ \hline
    \multicolumn{1}{c|}{Decision Tree} & AVT      & -        & \textbf{85.7}   & \textbf{85.7}      & \textbf{85.7}     & -             & 66.7          & 50.0          & 57.1          \\
    \multicolumn{1}{c|}{SVM}           & AVT      & -        & 42.8   & 60.0      & 50.0     & -             & 77.8          & 46.7          & 58.3          \\
    \multicolumn{1}{c|}{Topic Model}   & AVT      & -        & -      & -         & 70.0     & -             & -             & -             & 60.0          \\
    \multicolumn{1}{c|}{Random Forest} & T        & -        & 67     & 57        & 62       & -             & \textbf{89.0} & 40.0          & 55.0          \\
    \multicolumn{1}{c|}{TDCN}          & V        & \textbf{85.7}     & 83.3   & 77.0      & 80.0     & 66.0          & 64.3          & 45.0         & 53.0          \\ \hline
    \multicolumn{1}{c|}{TOAT}          & AT       & 78.8     & 83.3   & 66.7      & 74.1     & \textbf{73.9} & 78.6          & \textbf{55.0} & \textbf{64.7} \\ \hline
    \end{tabular}
    \end{table*}
	\subsection{Dataset and Implementation Details}
	We evaluate our method on a public dataset for multimodal depression detection named Distress Analysis Interview Corpus Wizard-of-Oz (DAIC-WOZ) \cite{daic}. In the interviews, an artificial intelligence named Ellie is used to ask participants some task-related questions and collect the recordings from the participants in the meanwhile.
	
	The DAIC-WOZ dataset includes three modalities: video, audio, and text. The visual features extracted from the recordings are preprocessed with the OpenFace toolkit \cite{baltrusaitis2016openface_11} for the sake of privacy protection. For the audio modality, besides the raw speech audio, the dataset also provides low-level features that are preprocessed by COVAREP toolbox \cite{degottex2014covarep}. For the textual modality, the text transcripts are provided. In our work, we use the raw speech audio and the text transcripts. The DAIC-WOZ dataset collects the interviews from 189 participants. Since a few samples (their IDs are 451, 458, and 480) have no question information, we ignore these samples in all our experiments. For the left 186 samples, the division in the experiments for training/validation/testing is 107/33/46. In addition, as the DAIC-WOZ dataset is somewhat imbalanced, we apply the strategy of random-oversampling \cite{gong2017topic_5,toto2021audibert} for the depressed subjects during the training process, aiming to alleviate the imbalance issue.

	We choose ten mostly-asked questions (\autoref{topic}) in our research ($N=10$). Each of them represents a topic. We implement TOAT with Pytorch \cite{paszke2019pytorch} and HuggingFace library \cite{wolf2020huggingfaces}. We use the AdamW optimizer \cite{loshchilov2019decoupled} and set the learning rate as 4e-6. We freeze the parameters of the feature encoder in wav2vec 2.0. The batch size of our model is set as 1. The depression detection is formulated as a classification task, where non-depressed is labeled as 0 (negative) and depressed is labeled as 1 (positive).

    \begin{table}[]
    \caption{The list of ten frequently asked questions and their sample numbers }
    \label{topic}
    \begin{tabular}{l|l|l}
    \hline
    Index & Question Description                                & Samples \\ \hline
    Q1    & How are you doing today?                            & 184     \\
    Q2    & How are you at controlling your temper?             & 176     \\
    Q3    & What'd you study at school?                         & 162     \\
    Q4    & Is there anything you regret?                       & 152     \\
    Q5    & Have you been diagnosed with depression?            & 167     \\
    Q6    & When was the last time you argued with someone      & 182     \\
          & and what was it about?                              &         \\
    Q7    & What advice would you give to yourself ten or       & 177     \\
          & twenty years ago?                                   &         \\
    Q8    & What are you most proud of in your life?            & 171     \\
    Q9    & When was the last time you felt really happy?       & 179     \\
    Q10   & How easy is it for you to get a good night's sleep? & 173     \\ \hline
    \end{tabular}
    \end{table}

	We evaluate the performance of TOAT with the metrics of accuracy, precision, recall, and F1-score. Their definitions are shown in Eq. (\ref{metrics}), in which TP, TN, FP, and FN denote True Positive, True Negative, False Positive, and False Negative, respectively. Specifically, "True" means a subject is predicted correctly, and "Positive" means its ground truth label is positive.
	
	\begin{equation}
	    \label{metrics}
		\begin{split}
			Accuracy= &\frac{TP+TN}{TP+TN+FP+FN}\\
			Recall= &\frac{TP}{TP+FN}\\
			Precision= &\frac{TP}{TP+FP}\\
			F1-score= &2\times\frac{Precision\times Recall}{Precision+Recall}
		\end{split}
	\end{equation}
	

    \subsection{Comparison with Other Methods}
    We compare our method with some methods to validate its effectiveness. These methods for comparison are chosen for that they do the classification task and report their results on the test set of DAIC-WOZ as we do:

    \noindent \textbf{Decision Tree} \cite{decision_tree} propose a method to fuse the high-level language information with low level audio and visual features.

    \noindent \textbf{SVM} \cite{2016avec_3} is a baseline method that uses a linear support vector machine proposed in AVEC 2016.

    \noindent \textbf{Topic Model} \cite{gong2017topic_5} performs topic modeling to extract topics and build multimodal features based on them. It then adopts a regression model to predict depression.

    \noindent \textbf{Random Forest} \cite{sun2017a} extracts textual features by specific topics such as sleep quality, PTSD/depression diagnostic.

    \noindent \textbf{TDCN} \cite{guo2022automatic} is a CNN-based method that extracts features from two visual cues and fuses them with the feature-wise attention.

	The above methods either use data of different modalities, or they extract features with different physical meanings from single-modality data. The quantitative comparison between our method and the is shown in \autoref{testset}. We have the following observations from the table. First, our model achieves the best performance on the test set. For example, TOAT surpasses the second-best Topic Model \cite{gong2017topic_5} in F1-score by 4.7\%. In addition, despite that Random Forest \cite{sun2017a} achieves the best Recall, its F1-score is still much lower than that of the TOAT model, due to its extremely low Precision score. Second, we can observe that the performance on the test set is generally lower than that on the validation set for all the models. The performance drop is attributed to the overfitting caused by the training on the small dataset. However, our model suffers the least from this issue among all the models, showing the usefulness of adopting the strategy of pre-training and fine tuning.

	\subsection{Ablation Study}
	
	\begin{table}[]
	\caption{Results of ablation studies on the DAIC-WOZ test set. $\alpha$ is the threshold of the topic attention. When it's set as 0, it means that there is no threshold. '-' represents there is no topic attention.}
	\label{ablation}
	\centering
    \begin{tabular}{cccccc}
    \hline
   \multicolumn{6}{c}{Multimodal}                                                                                                                          \\ \hline
\multicolumn{1}{c|}{Description}           & \multicolumn{1}{c|}{$\alpha$} & Accuracy      & Recall        & Precision     & F1-score      \\ \hline
\multicolumn{1}{c|}{\multirow{4}{*}{TOAT}} & \multicolumn{1}{c|}{\textbf{0.1}}          & \textbf{73.9} & \textbf{78.6}          & \textbf{55.0}          & \textbf{64.7}          \\
\multicolumn{1}{c|}{}                      & \multicolumn{1}{c|}{0}                     & 71.7          & 57.1          & 50.0          & 53.3          \\
\multicolumn{1}{c|}{}                      & \multicolumn{1}{c|}{0.2}                   & 69.5          & 0             & 0             & 0             \\
\multicolumn{1}{c|}{}                      & \multicolumn{1}{c|}{-}                     & 69.6          & 57.1          & 50.0          & 53.3          \\ \hline
\multicolumn{6}{c}{Unimodal}                                                                                                                            \\ \hline
\multicolumn{1}{c|}{\multirow{3}{*}{Text}} & \multicolumn{1}{c|}{0.1}                   & \textbf{73.9} & \textbf{50.0} & 53.9          & \textbf{51.9} \\
\multicolumn{1}{c|}{}                      & \multicolumn{1}{c|}{0}                     & 69.6          & \textbf{50.0} & 50.0          & 50.0          \\
\multicolumn{1}{c|}{}                      & \multicolumn{1}{c|}{-}                     & 67.4          & 42.9          & 46.2          & 44.4          \\ \hline
\multicolumn{1}{c|}{Audio}                 & \multicolumn{1}{c|}{-}                     & 71.7          & 42.9          & \textbf{54.6} & 48.0          \\ \hline
    \end{tabular}
    \end{table}

	In this subsection, we conduct several ablation studies to evaluate the effectiveness of the key elements in TOAT. The results are summarized in \autoref{ablation}.
	
	\begin{table*}[]
	\centering
	\caption{Two examples from the DAIC-WOZ test set. 'GT' represents 'Ground Truth'. Due to the limited space, superfluous words are omitted.}
	\label{example}
    \begin{tabular}{cc|c}
    \hline
    \multicolumn{1}{c|}{Index}   & Answer Description                                                                                                        & Score                \\ \hline
    \multicolumn{1}{c|}{Q1}      & uh i i feel i feel pretty good                                                                                   & 0               \\
    \multicolumn{1}{c|}{Q2}      & i think i'm pretty good at it                                                                                    & 0               \\
    \multicolumn{1}{c|}{Q3}      & uh i was i took two semestersof uh college and i was undecided so i was just taking general classes              & 0              \\
    \multicolumn{1}{c|}{Q4}      & no                                                                                                               & 0               \\
    \multicolumn{1}{c|}{Q5}      & \textbf{no}                                                                                                               & 0.1036               \\
    \multicolumn{1}{c|}{Q6}      & \textbf{\textless{}deep breath\textgreater wow uh yeah i can't remember the last time i had an argument with someone i don't know} & 0.1001                    \\
    \multicolumn{1}{c|}{Q7}      & ten or twenty years ago uh guess i would tell myself to stick with what what stick with what i love                        & 0                    \\
    \multicolumn{1}{c|}{Q8}      & \textbf{what am i most proud of like an event or an accomplishment}                                                                 & 0.1058                    \\
    \multicolumn{1}{c|}{Q9}      & uh xxx today when i talked when i uh talked to my friend earlier                                                 & 0               \\
    \multicolumn{1}{c|}{Q10}     & \textbf{very easy}                                                                                                                 & 0.1128                    \\ \hline
    \multicolumn{1}{c|}{GT}   & non-depressed                                                                                                             & {[}1, 0{]}           \\
    \multicolumn{1}{c|}{Prediction} & non-depressed                                                                                                             & {[}0.997, 0.003{]} \\ \hline\hline
    \multicolumn{1}{c|}{Q1}      & \textbf{i'm okey}                                                                                   & 0.1920               \\
    \multicolumn{1}{c|}{Q2}      & i try not to get angry um because i have a really bad temper and it um ...               & 0               \\
    \multicolumn{1}{c|}{Q3}      & i studied language um and um physics and um math              & 0              \\
    \multicolumn{1}{c|}{Q4}      & ugh \textless{}sigh\textgreater so much um my last marriage um not getting out when i should have and uh just i regret a lot          & 0               \\
    \multicolumn{1}{c|}{Q5}      & None                                                                                                              & 0               \\
    \multicolumn{1}{c|}{Q6}      & it was a couple days ago it was actually with my mother it was about actually she thought that ... & 0                    \\
    \multicolumn{1}{c|}{Q7}      & \textbf{\textless{}sigh\textgreater any situation that would've had a red flag just get out of it you know some things just don't work ...}         & 0.1927                    \\
    \multicolumn{1}{c|}{Q8}      & \textbf{i've had so many years of just bad just a lot of bad that there have been a long time since i've been proud ...}       & 0.2008                    \\
    \multicolumn{1}{c|}{Q9}      & \textbf{actuallyi'm kind of a couple days ago i was actually really happy but ... going back to feeling miserable so}                         & 0.1555               \\
    \multicolumn{1}{c|}{Q10}     & \textbf{it isn't um i tend to think a lot um about the things that stress me out um i've been through a lot of stuff ...}      & 0.1887                    \\ \hline
    \multicolumn{1}{c|}{GT}   & depressed                                                                                                             & {[}0, 1{]}           \\
    \multicolumn{1}{c|}{Prediction} & depressed                                                                                                             & {[}6e-4, 0.999{]} \\ \hline
    \end{tabular}
    \end{table*}

	\subsubsection{Topic Attention and Threshold}
	We evaluate the effect of the topic attention and the threshold first. In the first study, we remove the topic attention directly. Instead, we extend the subsets in the beginning to a long sentence, as the RoBERTa is able to process two sentences at most. From \autoref{ablation}, we can see the performance without the TA module has an obvious drop shows for both two versions of our model. For example, the F1-score drops from 64.7 (TOAT, $\alpha=0.1$) to 53.3 (w/o TA). As for the unimodal-based version, the F1-score drops from 51.9 (text, $\alpha=0.1$) to 44.4 (text w/o TA) by removing the TA module. The reasons are two-fold. First, as the tokenizer in \cite{roberta} receives a sentence whose length is 512 at most, we have to drop some information when we aggregate all questions and replies together. Second, the language model has difficulty in learning information from multiple Question+Answer pairs when we put them in one sentence, which are commonly separated by a $SEP$ token.

	The threshold $\alpha$ also has large impact on the performance. In our experiments, we achieve the best results by empirically setting $alpha$ as $1/N=0.1$. We observe that the overall performance is very sensitive to the threshold setting. On the one hand, if we set $alpha$ as 0, or keep originally learned weights for each topic in another word, the overall performance has an obvious drop in both two versions, which again shows the effectiveness of the topic attention mechanism. Surprisingly, when we increase the threshold $alpha$ into 0.2, the TOAT model totally fails to detect the depressed samples from the test set. In this threshold setting, we find that all the learned topic weights $\tilde{G_i}$ are suppressed as zero, as no $G_i^*$ is larger than $\alpha=0.2$. In this context, the so-called textual feature $\tilde{H}_t$ collapses into a zero vector, and therefore disables the whole model.


	\subsubsection{Unimodal Setting}
	In the ablation study, we also construct two incomplete versions of our model by removing the one of the branches. From the results in the Unimodal part of \autoref{ablation}, the audio branch itself achieves the F1-score of 48.0, while the textual branch itself obtains the F1-score of 51.9. The performance gap between the two incomplete versions and the full TOAT model ($\alpha=0.1$) demonstrate the necessity of combing multiple modalities. As the textual modality contains semantic information and the audio modality contains pronunciation and intonation information, they are highly complementary to each other, therefore forming the a more comprehensive representation.

	\subsubsection{Topic Analysis}
	
	\begin{figure}[htpb]
		\centering
		\includegraphics[width=\linewidth]{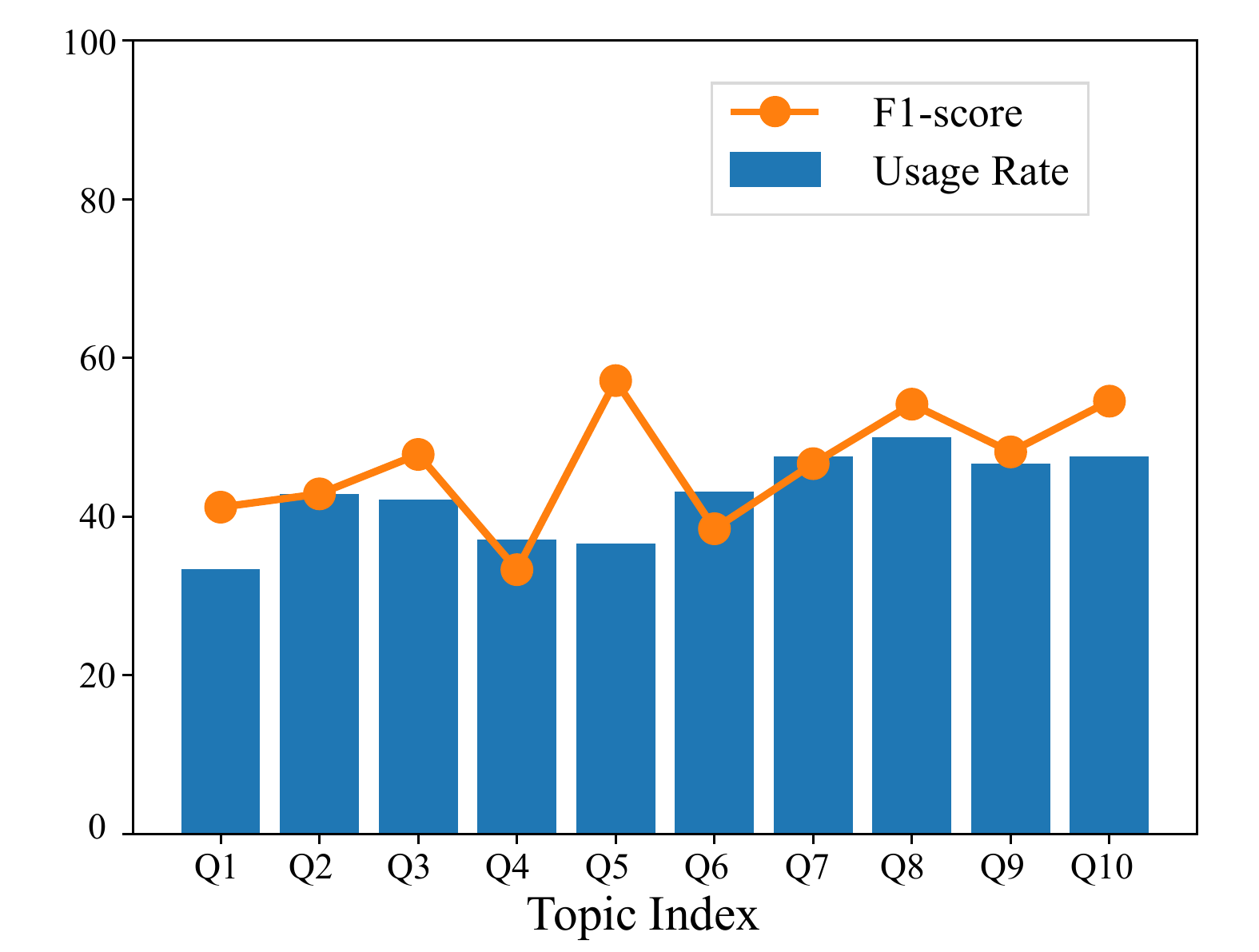}
		\caption{The topic-wise F1-scores and the usage rates ($\alpha=0.1$) based on the two experimental settings.}
		\label{topic_analysis}
	\end{figure}
	
	 We design another experiment to study the individual usefulness of each topic. We conduct the following experimental setting.  First, we reorganize the whole dataset into $N$ topic-wise subsets. For the $i$-th topic, we build the $i$-th subset by selecting the samples containing this topic, in which only the textual and audio data of this topic are retained. Then, we re-train the multiple versions of our model and test their performance based on these subsets. It is noted that we do not include the topic attention module in this experiment, as the data in these subsets only have one single topic. The F1-scores of the ten topics are shown in \autoref{topic_analysis}. In addition, we obtain the topic-related statistics from the experiment conducted on the whole DAIC-WOZ dataset. For each topic during the test, we accumulate its chosen time if the topic score of a sample is larger than the threshold $\alpha$. By dividing the chosen times by the number of samples (\autoref{topic}), we obtain the usage rates of all the topics, as shown in \autoref{topic_analysis}. From the figure, despite the F1-scores of these single-topic models stay at a low-level in general, they vary a lot across the different topics. In addition, we can see that the F1-scores of the topics are generally positively correlated with their usage rates, except the 5th topic. The results from the two experiments empirically present the different importance existed among the topics. In another word, if the performance of a certain topic is relatively better, it is more likely to be selected by the topic attention model. The exceptional case occurs in the 5th topic, i.e. 'Have you been diagnosed with depression?'. In this case, the trained model solely based on this topic is highly sensitive to its answer (yes or no). On the other hand, the results shown in \autoref{topic_analysis} demonstrate the necessity of utilizing multiple topics during the depression detection, as the result based on the whole dataset (F1-score of 64.7) is better than those of the ten models trained with their corresponding subsets.


    As is shown in \autoref{example}, we also provide two typical examples in the DAIC-WOZ test set. In the table, the replies of the topics chosen by the topic attention module ($G_i^*>=\alpha$) are highlighted in bold. From the answering details in \autoref{example}, we can see that the first participant is in a relatively good mood. He seldom argues about something, and feels easy to sleep. The second participant talks a lot, so we omit some words due to the limited space. We can see that this participant is not in a good mood. For example, she has difficulty in falling asleep, as she tends to think depressing things when she tries to sleep. She seldom feels happy, feeling miserable most time instead. As for these two cases, our model assigns the large weights to the replies that are highly related with the mental state inference, which is also generally in accordance with the subjective evaluation.

	\section{Conclusion}
	In this paper, we propose the TOAT model for the multimodal depression detection task. To alleviate the overfitting issue caused by small-dataset training, we construct the two-branch framework, in which two models pre-trained on large corpus are utilized, and fine-tunings are made based on the depression detection dataset. We also build the topic attention model for the textual modality branch, which fully explores the importance of each topic, and effectively enhances the detection accuracy.  Experimental results on the DAIC-WOZ dataset validates the effectiveness of our model. In the future, we plan to build a more comprehensive depression detection framework by introducing the other modalities such as visual cues.

	\textbf{Acknowledgements} This work was supported by National Key Research and Development Program (Grant No. 2019YFA0706200), the National Nature Science Foundation of China under Grant No. 62072152, 62172137, and the Fundamental Research Funds for the Central Universities under Grant No. PA2020GDKC0023.

\ifCLASSOPTIONcaptionsoff
  \newpage
\fi



%
    \bibliographystyle{IEEEtran}
    \bibliography{main}

\begin{thebibliography}{10}
\providecommand{\url}[1]{#1}
\csname url@samestyle\endcsname
\providecommand{\newblock}{\relax}
\providecommand{\bibinfo}[2]{#2}
\providecommand{\BIBentrySTDinterwordspacing}{\spaceskip=0pt\relax}
\providecommand{\BIBentryALTinterwordstretchfactor}{4}
\providecommand{\BIBentryALTinterwordspacing}{\spaceskip=\fontdimen2\font plus
\BIBentryALTinterwordstretchfactor\fontdimen3\font minus
  \fontdimen4\font\relax}
\providecommand{\BIBforeignlanguage}[2]{{%
\expandafter\ifx\csname l@#1\endcsname\relax
\typeout{** WARNING: IEEEtran.bst: No hyphenation pattern has been}%
\typeout{** loaded for the language `#1'. Using the pattern for}%
\typeout{** the default language instead.}%
\else
\language=\csname l@#1\endcsname
\fi
#2}}
\providecommand{\BIBdecl}{\relax}
\BIBdecl

\bibitem{2019Providing}
F.~Hao, G.~Pang, Y.~Wu, Z.~Pi, L.~Xia, and G.~Min, ``Providing appropriate
  social support to prevention of depression for highly anxious sufferers,''
  \emph{IEEE Transactions on Computational Social Systems}, pp. 1--9, 2019.

\bibitem{dep_refer1}
\BIBentryALTinterwordspacing
M.~Friedrich, ``\BIBforeignlanguage{en}{Depression {Is} the {Leading} {Cause}
  of {Disability} {Around} the {World}},''
  \emph{\BIBforeignlanguage{en}{JAMA}}, vol. 317, no.~15, p. 1517, 2017.
  [Online]. Available:
  \url{http://jama.jamanetwork.com/article.aspx?doi=10.1001/jama.2017.3826}
\BIBentrySTDinterwordspacing

\bibitem{kroenke2009the_14}
K.~{Kroenke}, T.~W. {Strine}, R.~L. {Spitzer}, J.~B. {Williams}, J.~T. {Berry},
  and A.~H. {Mokdad}, ``The phq-8 as a measure of current depression in the
  general population,'' \emph{Journal of Affective Disorders}, vol. 114, no.~1,
  pp. 163--173, 2009.

\bibitem{wang2008automated_1}
P.~{Wang}, F.~{Barrett}, E.~{Martin}, M.~{Milonova}, R.~E. {Gur}, R.~C. {Gur},
  C.~{Kohler}, and R.~{Verma}, ``Automated video-based facial expression
  analysis of neuropsychiatric disorders,'' \emph{Journal of Neuroscience
  Methods}, vol. 168, no.~1, pp. 224--238, 2008.

\bibitem{2021ta1}
M.~{Niu}, K.~{Chen}, Q.~{Chen}, and L.~{Yang}, ``Hcag: A hierarchical
  context-aware graph attention model for depression detection,'' in
  \emph{ICASSP 2021 - 2021 IEEE International Conference on Acoustics, Speech
  and Signal Processing (ICASSP)}, 2021.

\bibitem{hanai2018detecting_6}
T.~A. {Hanai}, M.~M. {Ghassemi}, and J.~R. {Glass}, ``Detecting depression with
  audio/text sequence modeling of interviews,'' in \emph{Proceedings of the
  Annual Conference of the International Speech Communication Association,
  INTERSPEECH}, 2018, pp. 1716--1720.

\bibitem{2019Detecting}
M.~Rohanian, J.~Hough, and M.~Purver, ``Detecting depression with word-level
  multimodal fusion,'' in \emph{Interspeech 2019}, 2019.

\bibitem{daic}
\BIBentryALTinterwordspacing
J.~Gratch, R.~Artstein, G.~Lucas, G.~Stratou, S.~Scherer, A.~Nazarian, R.~Wood,
  J.~Boberg, D.~DeVault, S.~Marsella, D.~Traum, S.~Rizzo, and L.-P. Morency,
  ``The distress analysis interview corpus of human and computer interviews,''
  in \emph{Proceedings of the Ninth International Conference on Language
  Resources and Evaluation ({LREC}'14)}.\hskip 1em plus 0.5em minus 0.4em\relax
  Reykjavik, Iceland: European Language Resources Association (ELRA), 2014, pp.
  3123--3128. [Online]. Available:
  \url{http://www.lrec-conf.org/proceedings/lrec2014/pdf/508_Paper.pdf}
\BIBentrySTDinterwordspacing

\bibitem{transformer}
\BIBentryALTinterwordspacing
A.~Vaswani, N.~Shazeer, N.~Parmar, J.~Uszkoreit, L.~Jones, A.~N. Gomez,
  L.~Kaiser, and I.~Polosukhin, ``Attention is all you need,'' in
  \emph{Advances in Neural Information Processing Systems 30: Annual Conference
  on Neural Information Processing Systems 2017, December 4-9, 2017, Long
  Beach, CA, {USA}}, I.~Guyon, U.~von Luxburg, S.~Bengio, H.~M. Wallach,
  R.~Fergus, S.~V.~N. Vishwanathan, and R.~Garnett, Eds., 2017, pp. 5998--6008.
  [Online]. Available:
  \url{https://proceedings.neurips.cc/paper/2017/hash/3f5ee243547dee91fbd053c1c4a845aa-Abstract.html}
\BIBentrySTDinterwordspacing

\bibitem{devlin2018bert}
\BIBentryALTinterwordspacing
J.~Devlin, M.-W. Chang, K.~Lee, and K.~Toutanova, ``{BERT}: Pre-training of
  deep bidirectional transformers for language understanding,'' in
  \emph{Proceedings of the 2019 Conference of the North {A}merican Chapter of
  the Association for Computational Linguistics: Human Language Technologies,
  Volume 1 (Long and Short Papers)}.\hskip 1em plus 0.5em minus 0.4em\relax
  Minneapolis, Minnesota: Association for Computational Linguistics, 2019, pp.
  4171--4186. [Online]. Available: \url{https://aclanthology.org/N19-1423}
\BIBentrySTDinterwordspacing

\bibitem{yang2019xlnet}
\BIBentryALTinterwordspacing
Z.~Yang, Z.~Dai, Y.~Yang, J.~G. Carbonell, R.~Salakhutdinov, and Q.~V. Le,
  ``Xlnet: Generalized autoregressive pretraining for language understanding,''
  in \emph{Advances in Neural Information Processing Systems 32: Annual
  Conference on Neural Information Processing Systems 2019, NeurIPS 2019,
  December 8-14, 2019, Vancouver, BC, Canada}, H.~M. Wallach, H.~Larochelle,
  A.~Beygelzimer, F.~d'Alch{\'{e}}{-}Buc, E.~B. Fox, and R.~Garnett, Eds.,
  2019, pp. 5754--5764. [Online]. Available:
  \url{https://proceedings.neurips.cc/paper/2019/hash/dc6a7e655d7e5840e66733e9ee67cc69-Abstract.html}
\BIBentrySTDinterwordspacing

\bibitem{raffel2020exploring}
C.~Raffel, N.~Shazeer, A.~Roberts, K.~Lee, S.~Narang, M.~Matena, Y.~Zhou,
  W.~Li, and P.~J. Liu, ``Exploring the limits of transfer learning with a
  unified text-to-text transformer,'' 2020.

\bibitem{zhou2021ibot}
J.~Zhou, C.~Wei, H.~Wang, W.~Shen, C.~Xie, A.~Yuille, and T.~Kong, ``ibot:
  Image bert pre-training with online tokenizer,'' 2021.

\bibitem{bao2021beit}
H.~Bao, L.~Dong, and F.~Wei, ``Beit: Bert pre-training of image transformers,''
  2021.

\bibitem{he2021masked}
K.~He, X.~Chen, S.~Xie, Y.~Li, P.~Dollár, and R.~Girshick, ``Masked
  autoencoders are scalable vision learners,'' 2021.

\bibitem{wav2vec2}
\BIBentryALTinterwordspacing
A.~Baevski, Y.~Zhou, A.~Mohamed, and M.~Auli, ``wav2vec 2.0: {A} framework for
  self-supervised learning of speech representations,'' in \emph{Advances in
  Neural Information Processing Systems 33: Annual Conference on Neural
  Information Processing Systems 2020, NeurIPS 2020, December 6-12, 2020,
  virtual}, H.~Larochelle, M.~Ranzato, R.~Hadsell, M.~Balcan, and H.~Lin, Eds.,
  2020. [Online]. Available:
  \url{https://proceedings.neurips.cc/paper/2020/hash/92d1e1eb1cd6f9fba3227870bb6d7f07-Abstract.html}
\BIBentrySTDinterwordspacing

\bibitem{baevski2020vq}
\BIBentryALTinterwordspacing
A.~Baevski, S.~Schneider, and M.~Auli, ``vq-wav2vec: Self-supervised learning
  of discrete speech representations,'' in \emph{8th International Conference
  on Learning Representations, {ICLR} 2020, Addis Ababa, Ethiopia, April 26-30,
  2020}.\hskip 1em plus 0.5em minus 0.4em\relax OpenReview.net, 2020. [Online].
  Available: \url{https://openreview.net/forum?id=rylwJxrYDS}
\BIBentrySTDinterwordspacing

\bibitem{schneider2019wav2vec}
S.~{Schneider}, A.~{Baevski}, R.~{Collobert}, and M.~{Auli}, ``wav2vec:
  Unsupervised pre-training for speech recognition.'' in \emph{Interspeech
  2019}, 2019, pp. 3465--3469.

\bibitem{roberta}
Y.~Liu, M.~Ott, N.~Goyal, J.~Du, M.~Joshi, D.~Chen, O.~Levy, M.~Lewis,
  L.~Zettlemoyer, and V.~Stoyanov, ``Roberta: A robustly optimized bert
  pretraining approach,'' 2019.

\bibitem{2016avec_3}
M.~{Valstar}, J.~{Gratch}, B.~{Schuller}, F.~{Ringeval}, D.~{Lalanne}, M.~T.
  {Torres}, S.~{Scherer}, G.~{Stratou}, R.~{Cowie}, and M.~{Pantic}, ``Avec
  2016: Depression, mood, and emotion recognition workshop and challenge,'' in
  \emph{Proceedings of the 6th International Workshop on Audio/Visual Emotion
  Challenge}, 2016, pp. 3--10.

\bibitem{decision_tree}
\BIBentryALTinterwordspacing
L.~Yang, D.~Jiang, L.~He, E.~Pei, M.~C. Oveneke, and H.~Sahli, ``Decision tree
  based depression classification from audio video and language information,''
  in \emph{Proceedings of the 6th International Workshop on Audio/Visual
  Emotion Challenge}, ser. AVEC '16.\hskip 1em plus 0.5em minus 0.4em\relax New
  York, NY, USA: Association for Computing Machinery, 2016, p. 89–96.
  [Online]. Available: \url{https://doi.org/10.1145/2988257.2988269}
\BIBentrySTDinterwordspacing

\bibitem{oureshi2021gender}
S.~A. Oureshi, G.~Dias, S.~Saha, and M.~Hasanuzzaman, ``Gender-aware estimation
  of depression severity level in a multimodal setting,'' in \emph{2021
  International Joint Conference on Neural Networks (IJCNN)}.\hskip 1em plus
  0.5em minus 0.4em\relax IEEE, 2021, pp. 1--8.

\bibitem{ray2019multi_7}
A.~{Ray}, S.~{Kumar}, R.~{Reddy}, P.~{Mukherjee}, and R.~{Garg}, ``Multi-level
  attention network using text, audio and video for depression prediction,'' in
  \emph{Proceedings of the 9th International on Audio/Visual Emotion Challenge
  and Workshop}, 2019, pp. 81--88.

\bibitem{makiuchi2019multimodal}
M.~R. {Makiuchi}, T.~{Warnita}, K.~{Uto}, and K.~{Shinoda}, ``Multimodal fusion
  of bert-cnn and gated cnn representations for depression detection,'' in
  \emph{Proceedings of the 9th International on Audio/Visual Emotion Challenge
  and Workshop}, 2019, pp. 55--63.

\bibitem{vgg}
\BIBentryALTinterwordspacing
K.~Simonyan and A.~Zisserman, ``Very deep convolutional networks for
  large-scale image recognition,'' in \emph{3rd International Conference on
  Learning Representations, {ICLR} 2015, San Diego, CA, USA, May 7-9, 2015,
  Conference Track Proceedings}, Y.~Bengio and Y.~LeCun, Eds., 2015. [Online].
  Available: \url{http://arxiv.org/abs/1409.1556}
\BIBentrySTDinterwordspacing

\bibitem{lstm}
S.~{Hochreiter} and J.~{Schmidhuber}, ``Long short-term memory,'' \emph{Neural
  Computation}, vol.~9, no.~8, pp. 1735--1780, 1997.

\bibitem{rohanian2019detecting}
M.~{Rohanian}, J.~{Hough}, and M.~{Purver}, ``Detecting depression with
  word-level multimodal fusion.'' in \emph{Interspeech 2019}, 2019, pp.
  1443--1447.

\bibitem{hong2010empirical}
L.~{Hong} and B.~D. {Davison}, ``Empirical study of topic modeling in
  twitter,'' in \emph{Proceedings of the First Workshop on Social Media
  Analytics}, 2010, pp. 80--88.

\bibitem{gong2017topic_5}
Y.~{Gong} and C.~{Poellabauer}, ``Topic modeling based multi-modal depression
  detection,'' in \emph{Proceedings of the 7th Annual Workshop on Audio/Visual
  Emotion Challenge}, 2017, pp. 69--76.

\bibitem{toto2021audibert}
E.~Toto, M.~Tlachac, and E.~A. Rundensteiner, ``Audibert: A deep transfer
  learning multimodal classification framework for depression screening,'' in
  \emph{Proceedings of the 30th ACM International Conference on Information \&
  Knowledge Management}, 2021, pp. 4145--4154.

\bibitem{librispeech}
\BIBentryALTinterwordspacing
V.~Panayotov, G.~Chen, D.~Povey, and S.~Khudanpur, ``Librispeech: An {ASR}
  corpus based on public domain audio books,'' in \emph{2015 {IEEE}
  International Conference on Acoustics, Speech and Signal Processing, {ICASSP}
  2015, South Brisbane, Queensland, Australia, April 19-24, 2015}.\hskip 1em
  plus 0.5em minus 0.4em\relax {IEEE}, 2015, pp. 5206--5210. [Online].
  Available: \url{https://doi.org/10.1109/ICASSP.2015.7178964}
\BIBentrySTDinterwordspacing

\bibitem{dai2021improving}
Z.~{Dai}, H.~{Zhou}, Q.~{Ba}, Y.~{Zhou}, L.~{Wang}, and G.~{Li}, ``Improving
  depression prediction using a novel feature selection algorithm coupled with
  context-aware analysis.'' \emph{Journal of Affective Disorders}, vol. 295,
  pp. 1040--1048, 2021.

\bibitem{niu2021hcag}
M.~{Niu}, K.~{Chen}, Q.~{Chen}, and L.~{Yang}, ``Hcag: A hierarchical
  context-aware graph attention model for depression detection,'' in
  \emph{ICASSP 2021 - 2021 IEEE International Conference on Acoustics, Speech
  and Signal Processing (ICASSP)}, 2021, pp. 4235--4239.

\bibitem{ba2016layer}
\BIBentryALTinterwordspacing
J.~L. Ba, J.~R. Kiros, and G.~E. Hinton, ``Layer normalization,'' \emph{arXiv
  preprint arXiv:1607.06450}, 2016. [Online]. Available:
  \url{https://arxiv.org/abs/1607.06450}
\BIBentrySTDinterwordspacing

\bibitem{baltrusaitis2016openface_11}
T.~{Baltrusaitis}, P.~{Robinson}, and L.-P. {Morency}, ``Openface: An open
  source facial behavior analysis toolkit,'' in \emph{2016 IEEE Winter
  Conference on Applications of Computer Vision (WACV)}, 2016, pp. 1--10.

\bibitem{degottex2014covarep}
G.~{Degottex}, J.~{Kane}, T.~{Drugman}, T.~{Raitio}, and S.~{Scherer},
  ``Covarep — a collaborative voice analysis repository for speech
  technologies,'' in \emph{Proceedings of IEEE International Conference on
  Acoustics, Speech and Signal Processing (ICASSP 2014)}, 2014, pp. 960--964.

\bibitem{paszke2019pytorch}
\BIBentryALTinterwordspacing
A.~Paszke, S.~Gross, F.~Massa, A.~Lerer, J.~Bradbury, G.~Chanan, T.~Killeen,
  Z.~Lin, N.~Gimelshein, L.~Antiga, A.~Desmaison, A.~K{\"{o}}pf, E.~Yang,
  Z.~DeVito, M.~Raison, A.~Tejani, S.~Chilamkurthy, B.~Steiner, L.~Fang,
  J.~Bai, and S.~Chintala, ``Pytorch: An imperative style, high-performance
  deep learning library,'' in \emph{Advances in Neural Information Processing
  Systems 32: Annual Conference on Neural Information Processing Systems 2019,
  NeurIPS 2019, December 8-14, 2019, Vancouver, BC, Canada}, H.~M. Wallach,
  H.~Larochelle, A.~Beygelzimer, F.~d'Alch{\'{e}}{-}Buc, E.~B. Fox, and
  R.~Garnett, Eds., 2019, pp. 8024--8035. [Online]. Available:
  \url{https://proceedings.neurips.cc/paper/2019/hash/bdbca288fee7f92f2bfa9f7012727740-Abstract.html}
\BIBentrySTDinterwordspacing

\bibitem{wolf2020huggingfaces}
T.~Wolf, L.~Debut, V.~Sanh, J.~Chaumond, C.~Delangue, A.~Moi, P.~Cistac,
  T.~Rault, R.~Louf, M.~Funtowicz, J.~Davison, S.~Shleifer, P.~von Platen,
  C.~Ma, Y.~Jernite, J.~Plu, C.~Xu, T.~L. Scao, S.~Gugger, M.~Drame, Q.~Lhoest,
  and A.~M. Rush, ``Huggingface's transformers: State-of-the-art natural
  language processing,'' 2020.

\bibitem{loshchilov2019decoupled}
\BIBentryALTinterwordspacing
I.~Loshchilov and F.~Hutter, ``Decoupled weight decay regularization,'' in
  \emph{7th International Conference on Learning Representations, {ICLR} 2019,
  New Orleans, LA, USA, May 6-9, 2019}.\hskip 1em plus 0.5em minus 0.4em\relax
  OpenReview.net, 2019. [Online]. Available:
  \url{https://openreview.net/forum?id=Bkg6RiCqY7}
\BIBentrySTDinterwordspacing

\bibitem{sun2017a}
B.~{Sun}, Y.~{Zhang}, J.~{He}, L.~{Yu}, Q.~{Xu}, D.~{Li}, and Z.~{Wang}, ``A
  random forest regression method with selected-text feature for depression
  assessment,'' in \emph{Proceedings of the 7th Annual Workshop on Audio/Visual
  Emotion Challenge}, 2017, pp. 61--68.

\bibitem{guo2022automatic}
Y.~Guo, C.~Zhu, S.~Hao, and R.~Hong, ``Automatic depression detection via
  learning and fusing features from visual cues,'' 2022.

\end{thebibliography}

%

\begin{IEEEbiography}[{\includegraphics[width=1in,height=1.25in,clip,keepaspectratio]{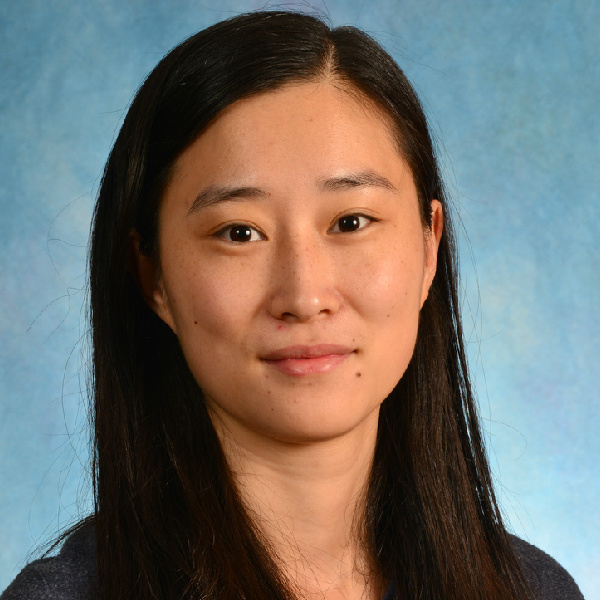}}]{Yanrong Guo}
Yanrong Guo is an associate professor at School of Computer and Information, Hefei University of Technology (HFUT). She is also with He is also with Key Laboratory of Knowledge Engineering with Big Data (Hefei University of technology), Ministry of Education. She received her Ph.D. degree at HFUT in 2013. She was a postdoc researcher at University of North Carolina at Chapel Hill (UNC) from 2013 to 2016. Her research interests include pattern recognition and medical image analysis.
\end{IEEEbiography}

\begin{IEEEbiography}[{\includegraphics[width=1in,height=1.25in,clip,keepaspectratio]{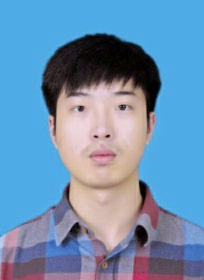}}]{Chenyang Zhu}
Chenyang Zhu received his B.E. from Hefei University of Technology, Hefei, in 2020. Now he is a master student at School of Computer Science and Information Engineering, Hefei University of Technology (HFUT). His current research interest is pattern recognition.
\end{IEEEbiography}


\begin{IEEEbiography}[{\includegraphics[width=1in,height=1.25in,clip,keepaspectratio]{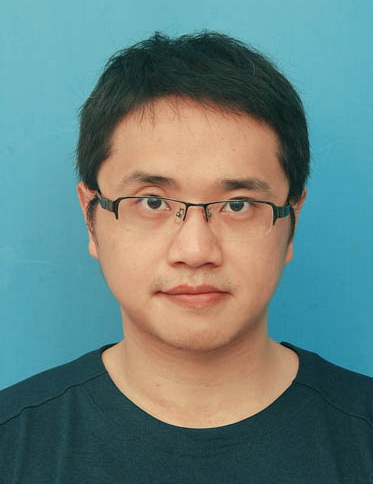}}]{Shijie Hao}
Shijie Hao is an associate professor at School of Computer Science and Information Engineering, Hefei University of Technology (HFUT). He is also with Key Laboratory of Knowledge Engineering with Big Data (Hefei University of technology), Ministry of Education. He received his Ph.D. degree at HFUT in 2012. His research interests include image processing and pattern recognition.
\end{IEEEbiography}

\begin{IEEEbiography}[{\includegraphics[width=1in,height=1.25in,clip,keepaspectratio]{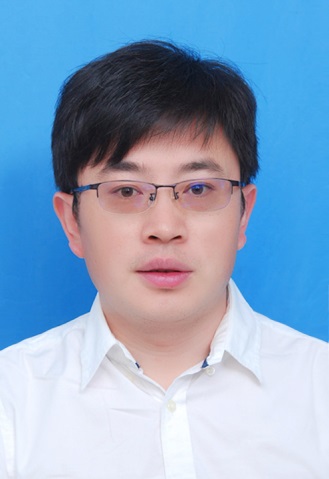}}]{Richang Hong}
Richang Hong received the Ph.D. degree from the University of Science and Technology of China, Hefei, China, in 2008. He was a Research Fellow of the School of Computing with the National University of Singapore, from 2008 to 2010. He is currently a Professor with the Hefei University of Technology, Hefei. He is also with Key Laboratory of Knowledge Engineering with Big Data (Hefei University of technology), Ministry of Education. He has coauthored over 70 publications in the areas of his research interests, which include multimedia content analysis and social media. He is a member of the ACM and the Executive Committee Member of the ACM SIGMM China Chapter. He was a recipient of the Best Paper Award from the ACM Multimedia 2010, the Best Paper Award from the ACM ICMR 2015, and the Honorable Mention of the IEEE Transactions on Multimedia Best Paper Award. He has served as the Technical Program Chair of the MMM 2016. He has served as an Associate Editor of IEEE Multimedia Magazine, Neural Processing Letter (Springer) Information Sciences (Elsevier) and Signal Processing (Elsevier).
\end{IEEEbiography}



\end{document}